\documentclass[runningheads]{llncs}

\usepackage[T1]{fontenc}
\usepackage{amssymb}

\usepackage{subcaption}    
\usepackage{float} 

\usepackage{multirow}
\usepackage[table,xcdraw]{xcolor}
\usepackage{booktabs}               
\usepackage{graphicx}               

\usepackage{amsmath} 

\usepackage{amssymb} 
\usepackage{graphicx}
\usepackage{placeins}

\begin{document}
\title{Degradation-Aware Blur-Segmentation \\ of Brain Tumor}

\author{
Yuchun Wang\and
Xiaosong Li\textsuperscript{*} \and
Gefei Liang \and
Yang Liu\
}
\authorrunning{Y. Wang et al.}
\institute{
School of Physics and Optoelectronic Engineering, Foshan University, China.
\email{2112455008@stu.fosu.edu.cn, lixiaosong@buaa.edu.cn}\\
\textsuperscript{*} Corresponding author
}
\maketitle             
\begin{abstract}
Multimodal 3D MRI brain tumor segmentation is a pivotal step in radiotherapy target delineation, surgical planning and post-treatment assessment. Existing methods often assume artifact-free MRI images. However, inevitable patient motion during scanning introduces artifacts and blur that degrade boundary and texture features, leading to poor segmentation performance. To bridge this gap, we introduce \textbf{D}egradation-\textbf{A}ware \textbf{B}lur-\textbf{Seg}mentation Net (DABSeg), a synchronous deblurring 3D multimodal MRI segmentation network that unifies blur removal and accurate segmentation. Specifically, we propose a feature-domain motion-deblurring stem to compensate for blur and rebalance intensity. Concurrently, the backbone network embeds a blur-aware cross-modal cross-attention module and multi-scale residual aggregation to yield effective modality complementarity. Notably, we optimize a joint loss that combines weighted Dice with a clear-reference reconstruction term, where imbalanced weights are applied to small targets to boost learning intensity and predictive stability for small lesions and border regions. Systematic comparisons and ablation experiments on the BraTS 2020 dataset under both clear and degenerative conditions consistently demonstrate that DABSeg surpasses state-of-the-art methods in tumor Dice score and boundary precision. These results validate the effectiveness of degenerative-aware cross-task collaborative learning in improving the robustness and clinical utility of multi-modal 3D brain tumor segmentation under realistic degenerative conditions. The source code is available at \url{https://github.com/YuchunWang24/DABSeg\_ICPR}.

\end{abstract}
\section{Introduction}
Brain tumors are space-occupying lesions that severely threaten life and health. Magnetic resonance imaging (MRI), as an important non-invasive tool for clinical diagnosis and therapeutic efficacy assessment~\cite{ref1}, can provide high-resolution soft-tissue contrast information. Within the multimodal MRI framework, T1-weighted (T1), contrast-enhanced T1-weighted (T1ce), T2-weighted (T2), and fluid-attenuated inversion recovery (FLAIR) characterize tumor enhancement, the core, and peritumoral oedema through different imaging mechanisms, providing complementary information for fine-grained segmentation. Tumors are typically partitioned into three key subregions: enhancing tumor (ET), tumor core (TC), and whole tumor (WT), and the segmentation results are critical for radiotherapy delineation, surgical planning, and prognostic assessment.

Existing studies on multimodal three-dimensional brain tumor segmentation have mainly focused on modality fusion and structural modeling. Liu~\cite{ref2} proposed KMD to improve generalization under incomplete multimodal settings by decomposing modality-shared and modality-specific features, while DC-Seg~\cite{ref3} enhanced robustness to missing modalities through decoupled contrastive learning and segmentation regularization. More recently, generative and diffusion-based methods have been explored. Qin \emph{et al.}~\cite{ref4} introduced Fourier-domain fusion with uncertainty-guided sampling, and Gao \emph{et al.}~\cite{ref5} employed axial attention to model long-range dependencies in a multi-scale diffusion framework. Despite these advances, existing methods generally assume ideally clear inputs and lack explicit modeling and unified optimization for motion blur degradation during acquisition.

However, under real imaging conditions, due to relatively long acquisition times, patients’ respiratory motion perturbs the k-space signal and forms motion artifacts during reconstruction, which commonly manifest as blurring, ringing, or ghosting, and are closely related to motion trajectories and sampling patterns~\cite{ref10}. Such degradations reduce image interpretability, increase the risk of repeated scans, and interfere with downstream automatic analysis and segmentation performance.

Research on motion artifacts has mainly focused on restoration and correction stages. IM-MoCo~\cite{ref6} performs self-supervised MRI motion correction using motion-guided implicit neural representations, while FDMC-Net~\cite{ref7} suppresses motion artifacts through frequency-domain decoupling and distillation. 
For downstream tasks, many existing studies still follow a two-stage paradigm, namely, ``restoration first, followed by downstream analysis''~\cite{ref18,ref19,ref20}. Sun \emph{et al.}~\cite{ref8} proposed BME-X, which first generates motion-corrected and denoised MRI images and then applies the enhanced images to downstream tasks such as tissue segmentation, registration, and diagnosis. Fujita  \emph{et al.}~\cite{ref9} incorporated motion-informed estimation into a deep-learning reconstruction pipeline for three-dimensional T1-weighted imaging, and evaluated its effectiveness through the reduction of downstream segmentation error.

In summary, existing multimodal three-dimensional brain tumor segmentation methods generally perform well under ideal inputs; however, clinical motion artifacts degrade boundary and texture contrast, leading to unstable segmentation and limited generalization. Current artifact removal studies predominantly follow a two-stage paradigm of ``restoration first, then segmentation,'' which makes it difficult to achieve joint optimization with segmentation objectives. Building upon this observation, we construct a deblurring–segmentation framework through the collaborative optimization of blur degradation compensation and blur-aware cross-modal fusion. \textbf{To the best of our knowledge, this is the first joint deblurring and segmentation framework for motion-blurred multimodal MRI}.

The main contributions of this paper are summarized as follows:
\begin{itemize}
  \item[(1)] We propose Degradation-Aware Blur-Segmentation Net (DABSeg), an end-to-end blur-aware cross-task collaborative learning deblurring–segmentation framework, which improves segmentation reliability under degraded multimodal three-dimensional inputs by jointly optimizing deblurring and segmentation objectives;
  \item[(2)] We design a Feature-Domain Motion Deblurring Stem. Without explicit motion kernel estimation or image-domain reconstruction, it learns residual correction and intensity re-normalization mappings in the voxel feature space, and is guided by the segmentation objective to obtain deblurring representations that are more aligned with downstream segmentation.

  \item[(3)] We perform systematic experiments under clear and blurred degradation conditions to quantify the blur impact on each tumor subregion and validate the robustness of the proposed method, providing reproducible evidence for segmentation under real-world acquisition conditions.
\end{itemize}

\section{Method}
\subsection{Overall Framework}

\begin{figure}  
\centering
\includegraphics[width=\textwidth]{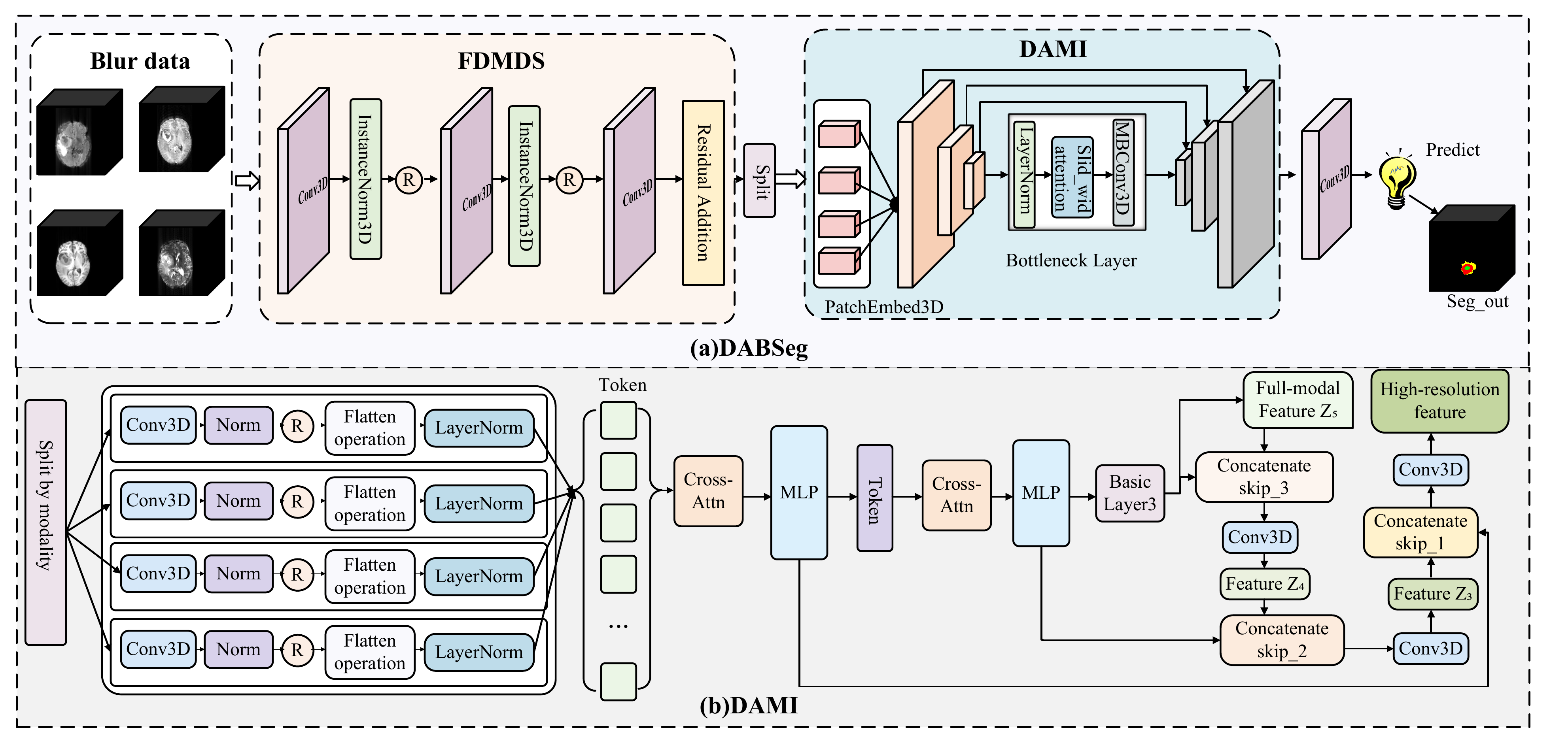}
\caption{(a) Overall framework of the proposed DABSeg. (b) Degradation-Aware Multi-Modal Interaction (DAMI).}
\label{fig1}
\end{figure}

In clinical practice, patient motion inevitably causes blur. ``Deblurring-first, segmentation-second'' methods optimize pixel-wise reconstruction, misaligned with segmentation-critical boundaries and subregions, producing clearer images but limited segmentation improvements. Based on these considerations, after constructing a motion-degraded dataset benchmark in the k-space domain, we propose DABSeg, a blur-aware end-to-end three-dimensional multimodal MRI deblurring–segmentation framework. Rather than first generating a ``perfectly corrected'' intermediate image and then performing segmentation, the overall design jointly models degradation compensation and tumor segmentation within a unified network, enabling the two tasks to form collaborative constraints in a shared feature space. The network takes four-modality motion-blurred volumetric data as input. Considering that motion degradation mainly manifests as the convolution of local blur kernels over voxel neighborhoods, while the global shape topology of tumors remains largely preserved before and after degradation, the blurred data are first processed by Feature-Domain Motion Deblurring Stem to perform lightweight voxel-level feature-domain deblurring and intensity renormalization, producing a `pseudo-clear' multimodal volume that carries degradation-compensation information. Subsequently, Degradation-Aware Multi-modal Interaction operates in a unified feature space to jointly model inter-modality complementary information and the distribution of degradation intensity through modality-related cross-attention and encoder–decoder-style multi-scale residual aggregation. 

At the network output, a lightweight 3D convolutional prediction head maps the fused voxel features to a three-channel tumor probability map $\mathrm{P}\in\mathrm{R}^{3\times \mathrm{D}\times \mathrm{H}\times \mathrm{W}}$, which is jointly optimized with the weighted Dice loss and the reconstruction term in the unified objective, thereby enabling end-to-end collaborative learning for deblurring and segmentation.

\subsection{Feature-Domain Motion Deblurring Stem(FDMDS)}

To address blur degradation introduced by patient motion in three-dimensional brain tumor MRI, we design a lightweight FDMDS as a front-end preceding the segmentation backbone. The core idea is not to explicitly estimate motion kernels in the physical acquisition process or in the k-space domain, but instead to directly learn a residual correction mapping in the voxel feature domain. As a feature-domain degradation compensation module, FDMDS is jointly optimized in an end-to-end manner together with the subsequent segmentation loss and the clear-reference volumetric reconstruction loss, thereby actively learning correction patterns that are beneficial for downstream tumor segmentation. In this way, the degraded four-modality volumetric data are transformed into a ``pseudo-clear'' feature volume that is more suitable for subsequent segmentation representation learning.
Let the multimodal volumetric data output by the FDMDS be denoted as $V_{\mathrm{blur}}\in\mathrm{R}^{\mathrm{C}\times \mathrm{D}\times \mathrm{H}\times \mathrm{W}}$, where the channel dimension $\mathrm{C}$ corresponds to the four modalities T1, T1ce, T2, and FLAIR, and $\mathrm{D}$, $\mathrm{H}$, and $\mathrm{W}$ denote the depth, height, and width of the volumetric data, respectively. The spatial dimensions of the input and output are kept unchanged throughout the process. Through cascaded convolutional operations and residual connections, the stem learns a mapping from $V_{\mathrm{blur}}$ to the deblurred feature volume $V$:
\begin{equation}
V = \mathrm{f}_{\mathrm{d}}\!\left(V_{\mathrm{blur}};\mathrm{\theta}_{\mathrm{d}}\right)
\in \mathrm{R}^{\mathrm{C}\times \mathrm{D}\times \mathrm{H}\times \mathrm{W}},
\end{equation}
where $\mathrm{f}_{\mathrm{d}}$ denotes the FDMDS and $\mathrm{\theta}_{\mathrm{d}}$ represents all learnable parameters. The specific computational process can be decomposed into three steps:

\noindent (1) Channel Expansion and Local 3D Receptive Field Enlargement.
First, a ConvBlock3D is applied to expand the input from $\mathrm{C}$ channels to $\mathrm{C}_{\mathrm{mid}}$ channels:
\begin{equation}
\mathrm{H}_{1}
=
\mathrm{\phi}_{1}\!\left(
\mathrm{N}_{1}\!\left(
\mathrm{W}_{1} * \mathrm{V}_{\mathrm{blur}} + \mathrm{b}_{1}
\right)
\right),
\quad
\mathrm{H}_{1} \in \mathrm{R}^{\mathrm{C}_{\mathrm{mid}} \times \mathrm{D} \times \mathrm{H} \times \mathrm{W}} .
\end{equation}
here, $\mathrm{C}_{\mathrm{mid}}$ denotes the intermediate feature dimension. $\mathrm{W}_{1}$, $\mathrm{b}_{1}$, $\mathrm{N}_{1}$, and $\mathrm{\phi}_{1}$ correspond to the convolution kernel, bias term, normalization operator, and activation function of the first layer, respectively. InstanceNorm3d is adopted as the normalization operator to mitigate global intensity scale variations across different patients and scanning protocols, enabling FDMDS to focus more on modeling local structural and textural degradation. Through channel expansion, the network increases channel capacity while preserving spatial resolution, providing richer representational capacity for encoding local texture and contrast cues related to motion blur.

\vspace{0.5em}
\noindent (2) Intermediate Feature Refinement.
Subsequently, another homogeneous ConvBlock3D is stacked to further refine the intermediate features:
\begin{equation}
\mathrm{H}_{2}
=
\mathrm{\phi}_{2}\!\left(
\mathrm{N}_{2}\!\left(
\mathrm{W}_{2} * \mathrm{V}_{\mathrm{blur}} + \mathrm{b}_{2}
\right)
\right) .
\end{equation}
here, $\mathrm{W}_{2}$, $\mathrm{b}_{2}$, $\mathrm{N}_{2}$, and $\mathrm{\phi}_{2}$ denote the convolution kernel, bias term, normalization operator, and activation function of the second layer, respectively. The combination of two consecutive $3 \times 3 \times 3$ convolutions results in an effective three-dimensional receptive field of approximately $5 \times 5 \times 5$. Compared with a single convolution, this configuration aggregates contextual information along the slice direction more effectively, thereby better capturing motion blur patterns that diffuse over time or along the slice axis.

\vspace{0.5em}
\noindent (3) Channel Reduction and Residual Correction.
The third layer employs a linear mapping consisting solely of a convolution to compress the features from $\mathrm{C}_{\mathrm{mid}}$ channels back to the original $\mathrm{C}$ channels:
\begin{equation}
\mathrm{R}
=
\mathrm{W}_{3} * \mathrm{H}_{2} + \mathrm{b}_{3},
\quad
\mathrm{R} \in \mathrm{R}^{\mathrm{C} \times \mathrm{D} \times \mathrm{H} \times \mathrm{W}} .
\end{equation}
here, $\mathrm{W}_{3} \in \mathrm{R}^{\mathrm{C} \times \mathrm{C}_{\mathrm{mid}} \times 3 \times 3 \times 3}$. Based on this mapping, an explicit residual connection is introduced by adding $\mathrm{R}$ to the original input $\mathrm{V}_{\mathrm{blur}}$. To complete motion blur compensation in the feature domain, a mild nonlinear transformation is applied to the residual result, yielding the final deblurred output:
\begin{equation}
\mathrm{V}
=
\mathrm{\phi}_{\mathrm{out}}\!\left(
\mathrm{R} + \mathrm{V}_{\mathrm{blur}}
\right) .
\end{equation}
here, $\mathrm{\phi}_{\mathrm{out}}$ denotes the LeakyReLU activation function.

\subsection{Degradation-Aware Multi-Modal Interaction (DAMI)}

After processing by FDMDS, the four-modality volumetric data have already undergone degradation compensation and intensity renormalization in the voxel space, yielding a ``pseudo-clear'' input that carries motion degradation priors. Based on this input, the segmentation backbone needs to address two key challenges simultaneously: first, how to fully exploit the complementary information among different modalities; and second, how to emphasize the representation of small-volume and boundary-blurred ET regions under realistic degradation conditions. To this end, the multimodal attention encoding and cross-scale feature integration in the backbone are unified and abstracted as a degradation-aware multimodal interaction trunk, termed DAMI, as illustrated in Fig.~1(b). DAMI is specifically designed to perform inter-modality interaction and multi-scale fusion within the degradation-compensated feature space.

\noindent (1) Degradation-Driven Multi-Modal Feature Embedding.

Let the output of FDMDS be denoted as $\mathrm{V} \in \mathrm{R}^{\mathrm{C} \times \mathrm{D} \times \mathrm{H} \times \mathrm{W}}$. To enable feasible modeling of different modalities within a unified feature space, DAMI first applies independent 3D convolutional embeddings to each modality, mapping the intensity space to a feature space with channel dimension $\mathrm{C}_{0}$. Let the modality set be denoted as $\mathrm{M} = \{1,2,3,4\}$. For any modality $\mathrm{m} \in \mathrm{M}$, the embedding is defined as
\begin{equation}
\mathrm{F}_{\mathrm{m}}^{0}
=
\mathrm{f}_{\mathrm{enc}}^{0}\!\left(
\mathrm{V}_{\mathrm{m}}
\right),
\quad
\mathrm{F}_{\mathrm{m}}^{0} \in \mathrm{R}^{\mathrm{C}_{0} \times \mathrm{D} \times \mathrm{H} \times \mathrm{W}} ,
\end{equation}
where $\mathrm{V}_{\mathrm{m}} \in \mathrm{R}^{1 \times \mathrm{D} \times \mathrm{H} \times \mathrm{W}}$ denotes the $\mathrm{m}$-th modality in the FDMDS output. The embedding function $\mathrm{f}_{\mathrm{enc}}^{0}(\cdot)$ consists of 3D convolution, normalization, and nonlinear activation, and its parameters are optimized end-to-end with the deblurring front-end through shared gradient propagation.

Subsequently, the three-dimensional feature maps are flattened into token sequences:
\begin{equation}
\mathrm{X}_{\mathrm{m}}^{0}
=
\mathrm{reshape}\!\left(
\mathrm{F}_{\mathrm{m}}^{0}
\right),
\quad
\mathrm{X}_{\mathrm{m}}^{0} \in \mathrm{R}^{\mathrm{N} \times \mathrm{C}_{0}},
\quad
\mathrm{N} = \mathrm{D} \times \mathrm{H} \times \mathrm{W} ,
\end{equation}
where $\mathrm{N}$ denotes the total number of voxels, and $\mathrm{X}_{\mathrm{m}}^{0}$ represents the initial token representation of modality $\mathrm{m}$ under the degradation-compensated coordinate system. Since FDMDS has already performed unified motion blur correction, tokens from different modalities are naturally aligned within the same spatial reference frame, providing the prerequisite for subsequent cross-modal attention modeling.

\noindent (2) Modality-Related Degradation-Aware Feature Interaction.

Simply stacking the four modalities along the channel dimension can increase the amount of information, but it often leads to redundancy, where less informative modalities may overwhelm modality-specific details that are critical for the representation of the enhancing tumor (ET). To address this issue, DAMI stacks $\mathrm{L}$ layers of modality-related interaction blocks at the encoding stage. Each layer explicitly models inter-modality dependencies through degradation-aware cross-attention.

For the $\mathrm{l}$-th layer, where $\mathrm{l} = 1, \ldots, \mathrm{L}$, the input consists of the token representations from the previous layer, denoted as $\mathrm{X}_{\mathrm{m}}^{\mathrm{l}-1} \in \mathrm{R}^{\mathrm{N} \times \mathrm{C}_{\mathrm{l}-1}}$. Linear projections are first applied to obtain the query, key, and value matrices:
\begin{equation}
\mathrm{Q}_{\mathrm{m}}^{\mathrm{l}} = \mathrm{X}_{\mathrm{m}}^{\mathrm{l}-1} \mathrm{W}_{\mathrm{Q}}^{\mathrm{l}},
\quad
\mathrm{K}_{\mathrm{m}}^{\mathrm{l}} = \mathrm{X}_{\mathrm{m}}^{\mathrm{l}-1} \mathrm{W}_{\mathrm{K}}^{\mathrm{l}},
\quad
\mathrm{V}_{\mathrm{m}}^{\mathrm{l}} = \mathrm{X}_{\mathrm{m}}^{\mathrm{l}-1} \mathrm{W}_{\mathrm{V}}^{\mathrm{l}},
\end{equation}
where $\mathrm{W}_{\mathrm{Q}}^{\mathrm{l}}$, $\mathrm{W}_{\mathrm{K}}^{\mathrm{l}}$, and $\mathrm{W}_{\mathrm{V}}^{\mathrm{l}} \in \mathrm{R}^{\mathrm{C}_{\mathrm{l}-1} \times \mathrm{d}_{\mathrm{l}}}$ are learnable projection matrices, and $\mathrm{d}_{\mathrm{l}}$ denotes the dimensionality of the attention subspace.

To explicitly capture modality complementarity, for a given modality $\mathrm{m}$, DAMI aggregates information not only from its own modality but also from the keys and values of all other modalities. The cross-modal aggregation is formulated as
\begin{equation}
\mathrm{Y}_{\mathrm{m}}^{\mathrm{l}}
=
\sum_{\mathrm{m}' \in \mathrm{M}}
\mathrm{Softmax}\!\left(
\frac{
\mathrm{Q}_{\mathrm{m}}^{\mathrm{l}}
\left(
\mathrm{K}_{\mathrm{m}'}^{\mathrm{l}}
\right)^{\top}
}{
\sqrt{\mathrm{d}_{\mathrm{l}}}
}
\right)
\mathrm{V}_{\mathrm{m}'}^{\mathrm{l}},
\end{equation}
where the Softmax operation is normalized along the token dimension. The resulting representation $\mathrm{Y}_{\mathrm{m}}^{\mathrm{l}} \in \mathrm{R}^{\mathrm{N} \times \mathrm{d}_{\mathrm{l}}}$ denotes the weighted fusion of all modalities for modality $\mathrm{m}$ at the $\mathrm{l}$-th layer within the degradation-compensated feature space.

In this manner, when a specific modality exhibits a low signal-to-noise ratio under the current motion degradation condition, its corresponding attention weights are automatically suppressed, and the network relies more heavily on modalities such as T2 and FLAIR, which are more sensitive to edema regions and lesion boundaries. The output of the cross-attention is further refined through residual connections and feed-forward networks.

\noindent (3) Multi-Scale Feature Aggregation and Residual Propagation.

Brain tumors exhibit both fine-grained boundaries and hierarchical shape structures. Performing modality interaction at a single scale is therefore insufficient to simultaneously capture local details and global topology. To address this issue, DAMI adopts a multi-scale encoder--decoder-style architecture along the spatial dimensions of the backbone. In this design, deeper features possess larger receptive fields, while shallower features preserve richer spatial details.

For clarity of presentation, the token representations updated at the $\mathrm{l}$-th layer are reshaped back into three-dimensional feature maps:
\begin{equation}
\mathrm{F}_{\mathrm{m}}^{\mathrm{l}}
=
\mathrm{reshape}^{-1}\!\left(
\mathrm{X}_{\mathrm{m}}^{\mathrm{l}}
\right),
\quad
\mathrm{F}_{\mathrm{m}}^{\mathrm{l}} \in \mathrm{R}^{\mathrm{C}_{\mathrm{l}} \times \mathrm{D}_{\mathrm{l}} \times \mathrm{H}_{\mathrm{l}} \times \mathrm{W}_{\mathrm{l}}},
\end{equation}
where $\mathrm{C}_{\mathrm{l}}$, $\mathrm{D}_{\mathrm{l}}$, $\mathrm{H}_{\mathrm{l}}$, and $\mathrm{W}_{\mathrm{l}}$ denote the channel number and spatial dimensions at the $\mathrm{l}$-th scale, respectively.

At the same scale, DAMI first aggregates features across the modality dimension. A compact formulation based on concatenation followed by a $1 \times 1 \times 1$ convolution is adopted:
\begin{equation}
\mathrm{F}_{\mathrm{fuse}}^{\mathrm{l}}
=
\Phi_{\mathrm{l}}\!\left(
\mathrm{F}_{1}^{\mathrm{l}}
\, \| \,
\mathrm{F}_{2}^{\mathrm{l}}
\, \| \,
\mathrm{F}_{3}^{\mathrm{l}}
\, \| \,
\mathrm{F}_{4}^{\mathrm{l}}
\right),
\end{equation}
where $\|$ denotes concatenation along the channel dimension, and $\Phi_{\mathrm{l}}(\cdot)$ represents a compression mapping composed of a $1 \times 1 \times 1$ convolution, normalization, and nonlinear activation. This operation performs cross-modal channel re-weighting with a limited number of parameters, further correcting the contribution of each modality at the feature level and complementing the attention weights introduced in the previous subsection.

Cross-scale aggregation is realized through upsampling and residual connections. Let $\mathrm{F}_{\mathrm{dec}}^{\mathrm{l}}$ denote the decoded feature at scale $\mathrm{l}$. The feature propagation to the next higher resolution scale is formulated as
\begin{equation}
\mathrm{F}_{\mathrm{dec}}^{\mathrm{l}-1}
=
\mathrm{G}_{\mathrm{l}-1}\!\left(
\mathrm{Up}\!\left(
\mathrm{F}_{\mathrm{dec}}^{\mathrm{l}}
\right)
\, \| \,
\mathrm{F}_{\mathrm{fuse}}^{\mathrm{l}-1}
\right),
\end{equation}
where $\mathrm{Up}(\cdot)$ denotes the spatial upsampling operator, $\|$ indicates feature concatenation, and $\mathrm{G}_{\mathrm{l}-1}(\cdot)$ represents the decoding convolutional block at scale $\mathrm{l}-1$. Through this top-down recursive update, deep semantic information is progressively propagated to higher-resolution scales, while boundary details from shallow features are fully integrated via residual paths.

Finally, the fused features obtained at the highest spatial resolution is denoted as
\begin{equation}
\mathrm{F}_{\mathrm{out}}
=
\mathrm{F}_{\mathrm{dec}}^{0}
\in
\mathrm{R}^{\mathrm{C}_{\mathrm{out}} \times \mathrm{D} \times \mathrm{H} \times \mathrm{W}} .
\end{equation}
this features integrates degradation-compensated multimodal contextual information and multi-scale structural representations, and is used as the unified segmentation feature input to the subsequent lightweight 3D convolutional prediction head, enabling voxel-wise classification of tumor subregions such as ET, TC, and WT. Through this degradation-aware multi-modal features interaction module, deblurring information, modality complementarity, and multi-scale structure are jointly incorporated into a unified feature learning framework without introducing additional complex branches.

\subsection{Joint Loss Function}

To simultaneously ensure deblurring quality and tumor segmentation accuracy under motion degradation, the network is optimized using a multi-term joint loss strategy. Overall, the training objective consists of a weighted Dice segmentation loss $\mathrm{L}_{\mathrm{seg}}$ and a deblurring reconstruction loss $\mathrm{L}_{\mathrm{rec}}$ based on clear reference data. The former directly supervises the three tumor probability maps predicted by DAMI, while the latter anchors the output of FDMDS to the corresponding clear volumetric data, thereby establishing consistency between degradation compensation and segmentation objective in the feature space. The joint objective is formulated as
\begin{equation}
\mathcal{L}
=
\mathrm{L}_{\mathrm{seg}}
+
\lambda_{\mathrm{rec}} \, \mathrm{L}_{\mathrm{rec}},
\end{equation}
where $\lambda_{\mathrm{rec}} > 0$ denotes the weight of the reconstruction loss. Based on extensive comparative experiments, $\lambda_{\mathrm{rec}}$ is fixed to $0.1$, which imposes a moderate constraint on the deblurring branch while preserving segmentation as the dominant optimization objective.

Considering that voxels belonging to the ET region are significantly fewer than those of TC and WT, and are also the most susceptible to being missed under degradation conditions, using uniform class weights throughout training would cause gradient updates to be dominated by WT and TC. This imbalance leads to insufficient representational capability for the ET subregion. To address this issue, after the network training stabilizes, explicit class weights $\mathrm{w}_{\mathrm{c}}$ are introduced for the three tumor categories. The weighted Dice segmentation loss is defined as follows:
\begin{equation}
\mathrm{Dice}_{\mathrm{c}}
=
\frac{1}{\mathrm{N}_{\mathrm{b}}}
\sum_{\mathrm{n}=1}^{\mathrm{N}_{\mathrm{b}}}
\mathrm{Dice}_{\mathrm{c}}^{\mathrm{n}},
\quad
\mathrm{c} \in \{\mathrm{ET}, \mathrm{TC}, \mathrm{WT}\},
\end{equation}
\begin{equation}
\mathrm{L}_{\mathrm{seg}}
=
1
-
\frac{
\sum_{\mathrm{c} \in \{\mathrm{ET}, \mathrm{TC}, \mathrm{WT}\}}
\mathrm{w}_{\mathrm{c}} \, \mathrm{Dice}_{\mathrm{c}}
}{
\sum_{\mathrm{c} \in \{\mathrm{ET}, \mathrm{TC}, \mathrm{WT}\}}
\mathrm{w}_{\mathrm{c}}
}.
\end{equation}
here, $\mathrm{N}_{\mathrm{b}}$ denotes the number of samples within a batch, $\mathrm{Dice}_{\mathrm{c}}^{\mathrm{n}}$ represents the Dice coefficient of the $\mathrm{n}$-th sample for class $\mathrm{c}$, and $\mathrm{Dice}_{\mathrm{c}}$ is the Dice score averaged over the batch dimension. The term $\mathrm{w}_{\mathrm{c}}$ denotes the class-specific weight.

\section{Experiments}

\subsection{Dataset}
The proposed DABSeg model is evaluated on the BraTS2020 dataset. Released through the Kaggle challenge platform, this dataset contains 369 annotated cases and an additional 125 unlabeled validation cases. Since ground-truth annotations are available only for the 369 labeled cases, they are split into training, validation, and test subsets with a ratio of 8:1:1, resulting in 295, 37, and 37 cases, respectively. Each case includes four MRI modalities (T1, T1ce, FLAIR, and T2) together with expert annotations. The original labels cover background, necrotic tissue, edema, and enhancing tumor regions. In this work, evaluation focuses on three clinically relevant tumor subregions, namely ET, WT, and TC, as shown in Fig.\ref{fig2}. All scans have a spatial resolution of $240 \times 240 \times 150$.

To better approximate motion artifacts under realistic acquisition conditions, we follow Shaw \textit{et al.}~\cite{ref10} to simulate motion artifacts on the four-modal 3D MRI data and construct a unified degraded dataset, denoted as BraTS2020\_S2, as shown in Fig.~\ref{fig3}. Performance is evaluated using the Dice similarity coefficient (Dice) and the 95\% Hausdorff distance (HD95), which measure volumetric overlap and boundary accuracy, respectively.

\subsection{Implementation Details}

In terms of implementation details, training is conducted on an NVIDIA RTX 3090 GPU. The deblurring front-end contains three 3D convolutional layers; the encoder adopts a three-stage hierarchical design with a depth configuration of $[2,2,2]$; and the decoder employs a five-stage progressive upsampling structure. The batch size is set to 1, the initial embedding scale is $4 \times 4 \times 4$, the embedding dimension is 32, and the total number of training epochs is 250. The initial learning rate is set to $1 \times 10^{-4}$ and is updated using a cosine annealing schedule. During training, randomly cropped 3D volumetric patches of size $128 \times 128 \times 128$ are used as input.    
\FloatBarrier
\begingroup
\setlength{\abovecaptionskip}{2pt}
\setlength{\belowcaptionskip}{0pt}

\begin{figure}[!t]
\centering

\includegraphics[width=\textwidth]{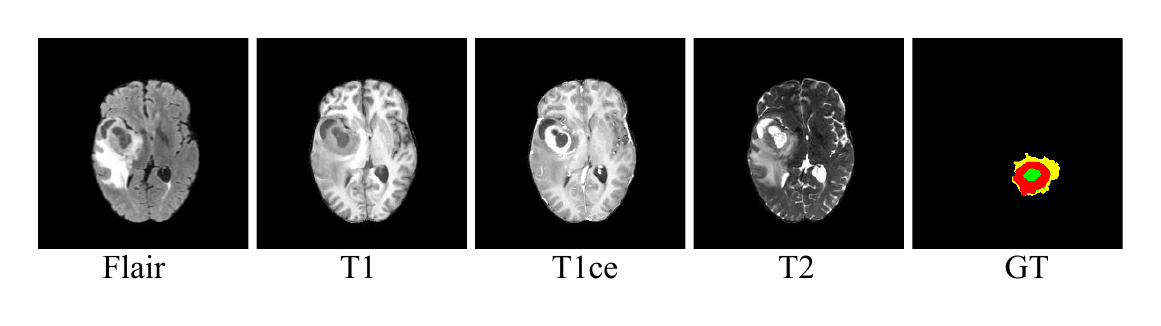}
\caption{Representative slices from 156 BraTS2020 patients, showing four MRI modalities and the corresponding ground-truth annotations. In the GT, red denotes ET, red and green indicate TC, and yellow, red, and green denote WT.}
\label{fig2}


\includegraphics[width=\textwidth]{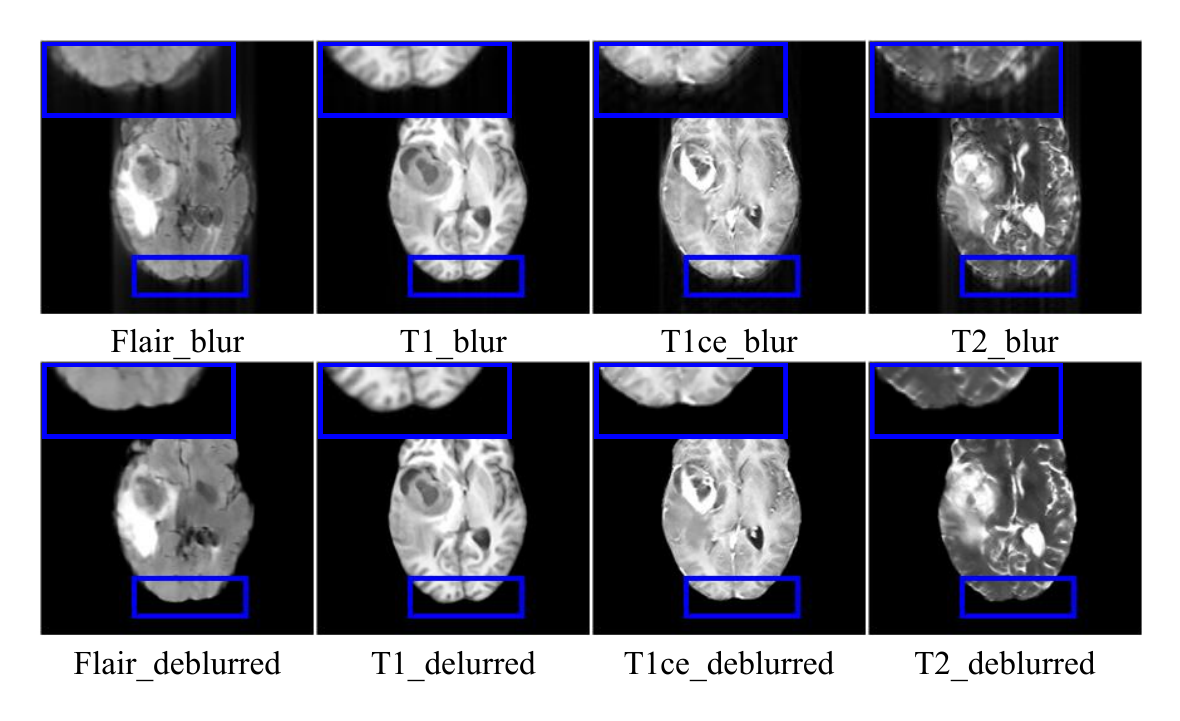}
\caption{Motion-blurred data BraTS2020\_S2 and deblurred data BraTS2020\_S2\_deblur.}
\label{fig3}

\end{figure}
\endgroup
\FloatBarrier

\subsection{Comparison with State-of-the-Art Methods}

To the best of our knowledge, no publicly available end-to-end restoration--segmentation methods are specifically designed for multimodal 3D brain tumor segmentation on the BraTS series datasets. For fair comparison, we adopt NAFNet~\cite{ref11} to deblur the BraTS2020\_S2 dataset, and all methods are evaluated on the resulting BraTS2020\_S2\_deblur dataset. As show in Fig.~\ref{fig3}. DABSeg is compared with Open\_brats~\cite{ref12}, CKD-TransBTS~\cite{ref13}, QT-Unet~\cite{ref14}, TDPC-NET~\cite{ref15}, DiffBTS~\cite{ref16} and HFF~\cite{ref17}.

\subsubsection{Quantitative Analysis}

As shown in Table\ref{tab1}, on the blurred BraTS2020\_S2 dataset, DABSeg achieves the best overall performance among all compared methods. Specifically, DABSeg obtains Dice scores of 86.88\%, 82.43\%, and 67.44\% for WT, TC, and ET, respectively, while the corresponding HD95 values are reduced to 8.71, 6.46, and 6.74, ranking as the best or second-best among all methods. Compared with CKD-TransBTS, DABSeg improves the Dice of the TC and ET regions by more than 38\% under blurred conditions. Although Open\_brats, QT-Unet, and TDPC-NET also achieve relatively good scores in the WT region, their performance on ET and TC is inferior and is accompanied by larger HD95 values, with average distances around 11. Notably, Table \ref{tab2} shows that on clean data, the overall performance gap among methods becomes smaller; nevertheless, DABSeg still maintains competitive Dice scores and lower HD95 values in the TC and ET regions, indicating that the proposed degradation-aware modeling does not sacrifice segmentation performance under clear conditions. Although DiffBTS achieves the best performance on WT in terms of HD95 with a score of 7.620, its Dice scores for all tumor subregions remain below 0.80. Overall, the proposed feature-domain motion deblurring module in DABSeg effectively alleviates the interference of degradation on multimodal representations, validating the effectiveness of the collaborative optimization strategy between blur compensation and segmentation.

\begin{figure} [!t]
\centering
\includegraphics[width=\textwidth]{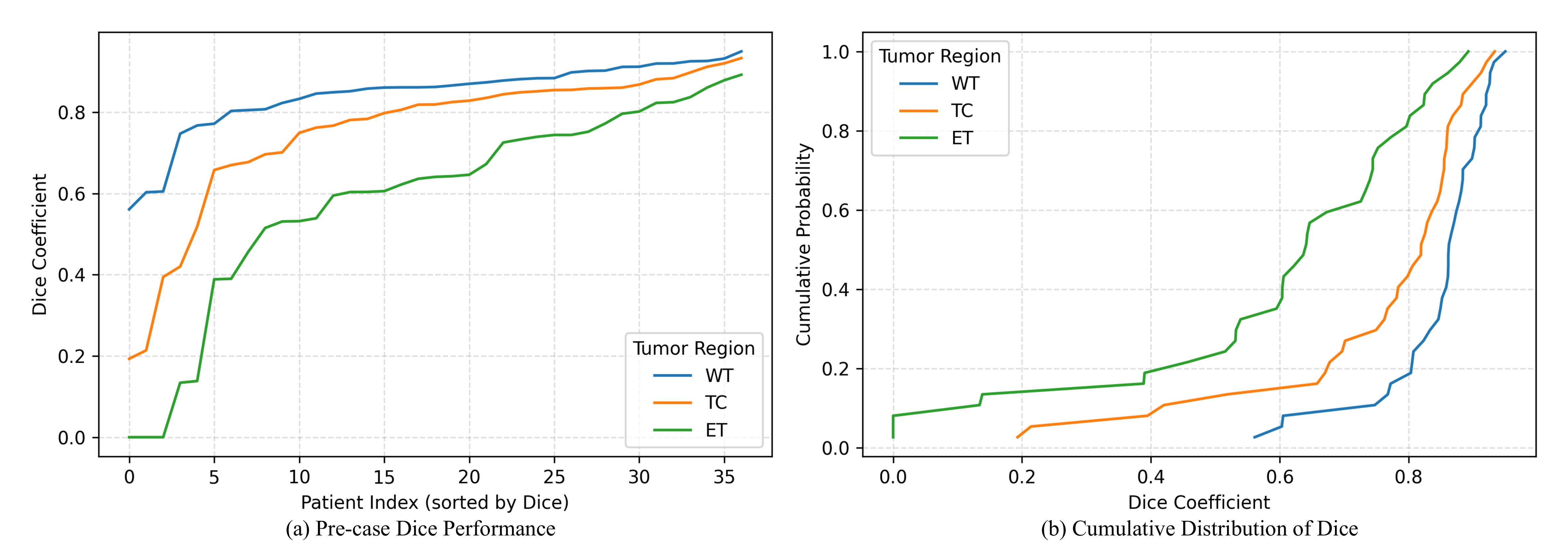}
\caption{Patient-level performance distribution of DABSeg on the BraTS2020\_S2 test set. (a) Patient-level performance curves sorted by Dice score; (b) the corresponding cumulative distribution function.}
\label{fig4}
\end{figure}

Fig.\ref{fig4}. shows the patient-level performance distribution of DABSeg on the BraTS2020\_S2 test set. Most cases exhibit high Dice scores for WT, TC, and ET. As shown in Fig.~4(a), the Dice-sorted curves rise sharply in the high-score range, while the cumulative distribution in Fig.~4(b) rapidly approaches 1 within the 0.8--0.9 interval. These results indicate consistent patient-level performance and further support the robustness of DABSeg.

\begin{table}[!t]
\centering
\caption{Comparison of segmentation metrics of different methods on the BraTS2020\_S2 test set. Dice values are reported as mean $\pm$ std across test cases}\label{tab1}
\begin{tabular}{lllllll}
\hline
\multicolumn{1}{c}{}                          & \multicolumn{2}{c}{WT}                                       & \multicolumn{2}{c}{TC}                                       & \multicolumn{2}{c}{ET}                                       \\
\multicolumn{1}{c}{\multirow{-2}{*}{Methods}} & Dice                          & HD95                           & Dice                          & HD95                           & Dice                          & HD95                           \\ \hline
0pen\_brats(2020)                             & 84.434 $\pm$ 14.792          & 13.526                         & 76.978 $\pm$ 12.452          & 9.484                          & 56.918 $\pm$ 28.546          & 29.177                         \\
CKD-TransBTS(2023)                            & 71.425 $\pm$ 8.807           & 45.147                         & 50.281 $\pm$ 28.803         & 41.410                         & 42.988 $\pm$ 21.876         & 56.984                         \\
QT-Unet(2024)                                 & 85.066 $\pm$ 8.183           & 7.643                          & 73.870 $\pm$ 13.681         & 8.353                          & 62.120 $\pm$ 24.811         & 8.547                          \\
TDPC-NET(2024)                                & 84.386 $\pm$ 9.705           & 10.585                         & 75.282 $\pm$ 20.361         & 10.268                         & 58.624 $\pm$ 25.536         & 10.215                         \\
DiffBTS(2025)                                 & 77.310 $\pm$ 19.858          & 7.620                          & 67.200 $\pm$ 24.184         & 8.320                          & 54.690 $\pm$ 22.012         & 11.340                         \\
HFF(2025)                                     & 85.700 $\pm$ 10.805          & {\color[HTML]{FE0000} 7.014}   & 80.750 $\pm$ 14.632         & 6.819                          & 62.910 $\pm$ 25.331         & 8.783                          \\
Ours                                          & {\color[HTML]{FE0000} 86.881} $\pm$ 7.164 & 8.708           & {\color[HTML]{FE0000} 82.433} $\pm$ 11.185 & {\color[HTML]{FE0000} 6.458} & {\color[HTML]{FE0000} 67.444} $\pm$ 23.950 & {\color[HTML]{FE0000} 6.735} \\ \hline
\end{tabular}
\end{table}

\begin{table}[]
\centering
\caption{Comparison of segmentation metrics of different methods on the BraTS2020 test set. Red is best, Blue is second.}\label{tab2}
\begin{tabular}{lllllll}
\hline
\multicolumn{1}{c}{}                          & \multicolumn{2}{c}{WT}                                       & \multicolumn{2}{c}{TC}                                       & \multicolumn{2}{c}{ET}                                       \\
\multicolumn{1}{c}{\multirow{-2}{*}{Methods}} & Dice                          & HD95                           & Dice                          & HD95                           & Dice                          & HD95                           \\ \hline
Open\_brats(2020)                             & 88.595                        & 19.549                       & 84.273                        & 6.667                        & {\color[HTML]{FE0000} 78.507} & 20.361                       \\
CKD-TransBTS(2023)                            & 87.418                        & 8.140                        & 82.394                        & 6.698                        & 74.975                        & 8.110                        \\
QT-Unet(2024)                                 & {\color[HTML]{FE0000} 91.994} & 5.584                        & {\color[HTML]{3531FF} 85.110} & {\color[HTML]{3531FF} 5.998} & 75.359                        & 6.961                        \\
TDPC-NET(2024)                                & 90.010                        & {\color[HTML]{FE0000} 4.490} & 83.430                        & 10.370                       & {\color[HTML]{3531FF} 78.180} & 30.560                       \\
DiffBTS(2025)                                 & 85.390                        & {\color[HTML]{3531FF} 5.040} & 75.180                        & 7.880                         & 62.730                        & {\color[HTML]{3531FF} 7.250}  \\
HFF(2025)                                     & 84.050                        & 8.942                        & 79.495                        & 7.123                        & 61.730                        & 8.203                        \\
Ours                                          & {\color[HTML]{3531FF} 90.374} & 7.040                        & {\color[HTML]{FE0000} 85.452} & {\color[HTML]{FE0000} 5.931} & 76.975                        & {\color[HTML]{FE0000} 6.883} \\ \hline
\end{tabular}
\end{table}

\subsubsection{Qualitative Analysis}
\begin{figure}[!t]
\centering
\includegraphics[width=\textwidth]{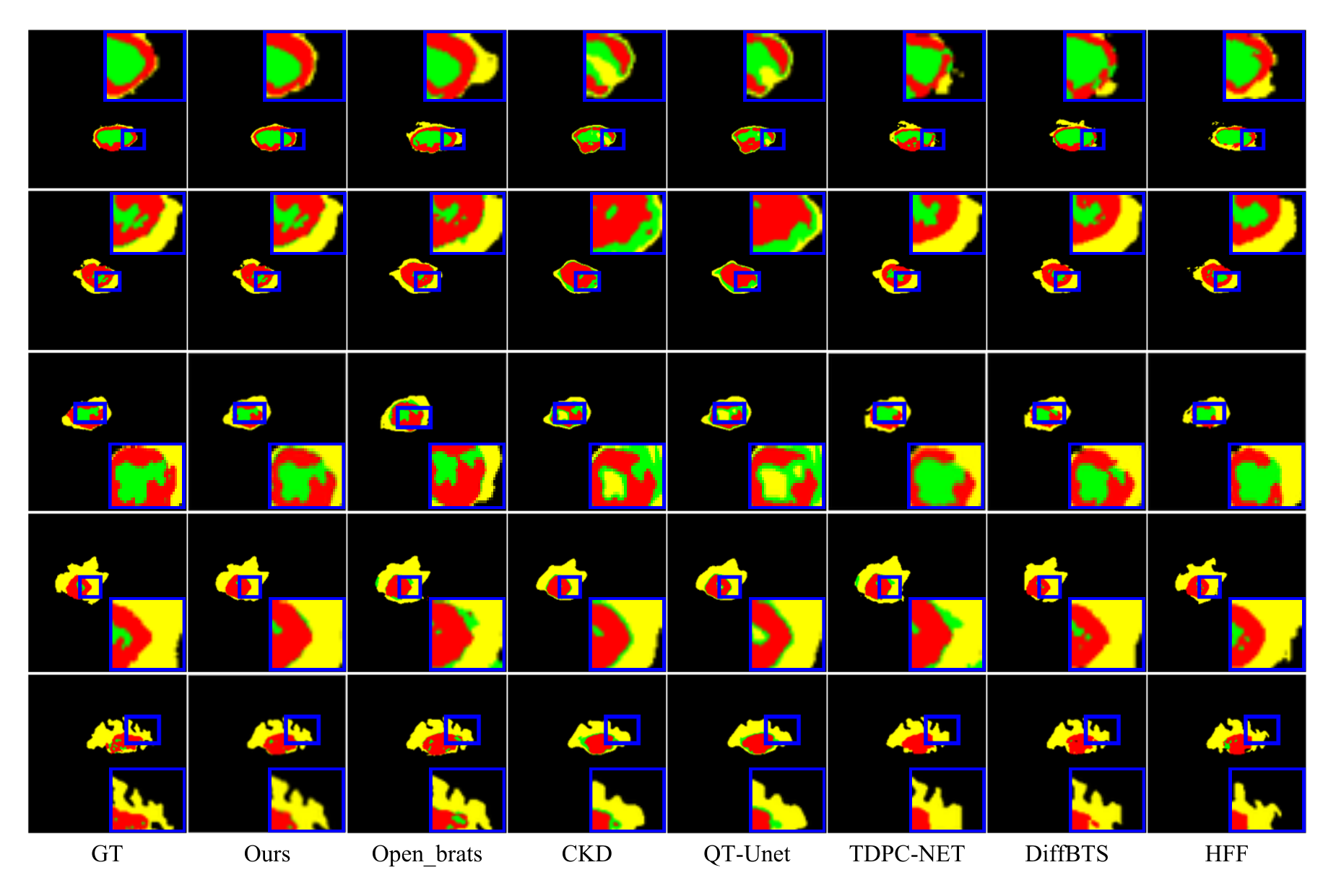}
\caption{Comparison of segmentation results of different methods on BraTS2020\_S2. 
}
\label{fig5}
\end{figure}

 To provide an intuitive comparison among methods, we present two qualitative visualization examples in Fig.\ref{fig5}. DABSeg shows the best morphological agreement with the GT: the WT outer boundary is well aligned and smoother, the TC extent is more stable, and the ET region preserves more continuous location and shape, with fewer jagged boundaries and class confusion. In contrast, CKD and QT-Unet are more prone to subregion hierarchy instability and local mis-segmentation under degraded inputs, manifested as boundary fluctuations and irregular internal structures; notably, ET is often over-segmented and accompanied by a thinning WT outer ring. 0pen\_brats and TDPC-NET can cover the major WT area, but still exhibit local leakage or unstable internal region allocation, and the shape consistency of ET/TC is weaker than that of Ours. DiffBTS provides relatively complete overall coverage, yet boundary expansion and over-smoothed details are observed in some samples; for HFF, although the contours appear smoother, fine-grained boundaries and subregion details remain insufficiently preserved.

Overall, the proposed end-to-end collaborative framework enhances the spatial consistency of ET and TC while preserving the WT contour, thereby improving the reliability and interpretability of segmentation under motion-degraded conditions.

\subsection{Ablation Study}

To validate the effectiveness of the degradation suppression branch and the cross-task joint optimization strategy in the end-to-end framework, we construct four model variants under the same training and evaluation protocol. As shown in Table \ref{tab3}, compared with the segmentation-only DAMI baseline, introducing FDMDS with end-to-end joint optimization leads to an overall Dice improvement of approximately 18-21 percentage points for WT, TC, and ET, together with a substantial reduction in HD95. This indicates that the collaborative modeling of degradation suppression and segmentation within a unified framework effectively enhances structural discrimination and boundary stability under motion degradation. Further adjusting $\lambda_{\mathrm{rec}}$ from 0.2 to 0.1 yields an additional Dice improvement of about 1 percentage point across all subregions, suggesting that moderately relaxing the reconstruction constraint facilitates more segmentation-oriented joint representations. Finally, incorporating ET\_weight further reduces the HD95 of the ET region while maintaining a comparable Dice score, indicating improved boundary localization accuracy, which is consistent with the small volume and boundary-sensitive characteristics of ET.

\begin{table}[!t]
\centering
\caption{Ablation results of DABSeg on the BraTS2020\_S2 test set.}\label{tab3}
\begin{tabular}{lllllll}
\hline
\multicolumn{1}{c}{}                          & \multicolumn{2}{c}{WT}                                       & \multicolumn{2}{c}{TC}                                       & \multicolumn{2}{c}{ET}                                       \\
\multicolumn{1}{c}{\multirow{-2}{*}{Methods}} & Dice                          & HD95                           & Dice                          & HD95                           & Dice                          & HD95\\ \hline
DAMI& 67.621                        & 43.420                       & 56.014                        & 43.466                       & 52.790                        & 44.919                       \\
DAMI+FDMDS($\lambda_{\mathrm{rec}}=0.2$)& 85.627                        & 10.636                       & 80.554                        & 7.273                        & 66.094                        & 7.696                        \\
DAMI+FDMDS($\lambda_{\mathrm{rec}}=0.1$)& 86.629                        & 8.754                        & 81.671                        & 6.648                        & {\color[HTML]{FF0000} 67.444} & 7.378                        \\
DAMI+FDMDS+ET\_weight($\lambda_{\mathrm{rec}}=0.1$)& {\color[HTML]{FF0000} 86.881} & {\color[HTML]{FF0000} 8.708} & {\color[HTML]{FF0000} 82.433} & {\color[HTML]{FF0000} 6.458} & {\color[HTML]{FF0000} 67.444} & {\color[HTML]{FF0000} 6.735} \\ \hline
\end{tabular}
\end{table}

\section{Conclusion}

This paper addresses motion blur degradation problem and proposes a blur-aware end-to-end three-dimensional multimodal brain tumor segmentation framework, termed DABSeg. Unlike existing methods that assume clear inputs or adopt stage-wise pipelines, DABSeg explicitly models motion degradation at the feature level and unifies blur compensation, multimodal complementary modeling, and segmentation objectives within a single optimization framework. This design effectively mitigates boundary ambiguity and small-lesion miss detection under degraded imaging conditions. Extensive experiments demonstrate that DABSeg consistently achieves superior segmentation accuracy and boundary stability across key tumor subregions, including WT, TC, and ET, validating the effectiveness of the feature-domain blur compensation branch, the blur-aware multimodal interaction mechanism, and the joint optimization strategy.

Future work will focus on extending degradation-aware modeling to more complex imaging scenarios, such as diverse acquisition artifacts and cross-center data distribution shifts, to further enhance the clinical applicability and deployment robustness of the proposed framework.

\section{Acknowledgement}
This work was supported by the National Natural Science Foundation of China (No. 62201149), and the Basic and Applied Basic Research of Guangdong Province (No. 2023A1515140077)

%
%
%

\bibliographystyle{splncs04}
\bibliography{references}

\end{document}